\documentclass[aps,pre,onecolumn,showpacs,a4paper]{revtex4}
\usepackage{graphicx} 
\usepackage{amsmath}
\usepackage{amssymb}
\usepackage{amscd}

\begin{document}

\title{Influence of the Noise Spectrum on
Stochastic Acceleration}

\author{K. Mallick}
 \affiliation{Service de Physique Th\'eorique, CEA Saclay, 91191
  Gif-sur-Yvette, France}
\date{\today}

\begin{abstract}
    We use an effective  Markovian description
 to   study the long-time behaviour of a nonlinear second order Langevin equation
 with Gaussian noise. When dissipation is neglected, the energy of the system grows
 as with time  a power-law  with an anomalous scaling exponent that depends
 both on the confining potential and on the high frequency distribution of the noise.
 The asymptotic expression of the Probability Distribution Function in phase space is
 calculated analytically. The results are extended to the case where small
   dissipative effects are taken into account.
  \end{abstract}
\maketitle 

 The influence of a   random perturbation on 
  a dynamical  system is a problem 
  of interest in various fields of  science and engineering
 \cite{arnold,crauel,lefever,toral,anishchenko, strato}. 
 The  first example of  a  differential
 equation with  stochastic terms
 appeared  in Langevin's study of Brownian motion \cite{vankampen,gardiner}: 
 Langevin modelized  the action  of the solvant molecules on the
 Brownian particle as the sum of a deterministic viscous
 friction, proportional to  the velocity of the  Brownian particle,  
 and of   a random force of auto-correlation
 proportional to the temperature of the bath.   Since then, it has been 
 customary to add in the dynamical equations
 some  phenomenological stochastic terms  that describe random
 environmental loadings (e.g. the influence of a turbulent wind on a suspension
 bridge, or  the study of random parametric vibration of helicopter rotor blades
 in atmospheric turbulent flow \cite{RobertsSpanos,ibrahim}). Of particular interest is the
 determination  of the energy flow into the system from external sources when
 the characteristic time  of the parameters variations matches  one of the natural
  frequencies of the system:  parametric resonance then occurs and 
 the rate of increase of the  amplitude is generally exponential  leading 
 to an instability. The growth of the response is limited by various nonlinear
 effects. 

   Several methods have been developed to study random  parametric  
 vibrations \cite{ibrahim}. One of the most efficient
 techniques  is the averaging principle  developed by Bogoliubov and Mitropol'skii
 for deterministic nonlinear vibrations \cite{bogoliubov}  where rapidly
 fluctuating circular coordinates are averaged out  leading to a set of effective
 dynamical equations for slow  variables.
 This method  was  extended to
 stochastic systems by Stratonovich \cite{strato} and put on a rigorous
 mathematical basis  by Khas'minskii \cite{khamin}  and by
  Papanicolaou and Kohler \cite{papanico}. Since then,  
 stochastic averaging  has become  a  powerful method  \cite{zhu1,zhu2} (for a recent 
 review see \cite{zhuRev} and references therein). 
 
 In a series of recent works \cite{philkir1,philkir2,philkir3,kolkata},
 we studied 
 the long time behaviour of  the  nonlinear oscillator subject to  parametric noise.
 We showed that,  in the absence of dissipation,  the  nonlinear terms
 in the potential stiffness inhibit  the exponential growth of the amplitude.
  The observables of the system (the
 amplitude, the momentum, the energy) rather  display power-law scalings  
 with anomalous diffusion exponents. When the parametric  noise is a Gaussian 
 white noise ({\it i.e.},  it has a vanishingly small correlation-time), the 
 averaging method, applied to the energy envelope  \cite{strato,lindcol,ibrahim,daffert,roberts},
 allows  us to calculate analytically the time-asymptotic probability
 distribution  function (PDF) of the system in phase space \cite{philkir1}; knowing this 
  PDF, the scaling exponents and the corresponding prefactors  are readily deduced. 
  However, for colored
 noise   a competition between conflicting time scales occurs.
  In fact, the nonlinear
 oscillator has an amplitude-dependent intrinsic frequency that increases
   with  the amplitude. And as  the oscillator absorbs energy from its
 environment, its amplitude grows  and  at a certain stage, the  intrinsic period 
  becomes  smaller than the correlation time of the noise.
  This corresponds to a cross-over regime 
  at  which the  correlation time  of the noise ceases to be the smallest time-scale
 in the system: the scaling laws that govern the growth  of  amplitude,  momentum
 and  energy differ from  those calculated for a white noise. Thus, 
  when the amplitude of the
 oscillator is small its intrinsic period is large and the noise appears as  if it were
 white and white noise exponents prevail; but at large amplitudes, the scaling regime
  changes and  new exponents appear.  Because of these conflicting time scales, the
 averaging   technique is  difficult to implement  for  a colored noise:  at lowest order
  the noise itself is averaged out and the energy transfer stops at the cross-over time.
 Therefore one has to perform  averaging at higher orders.
 When the colored noise  is an Ornstein-Uhlenbeck process
  the calculations can be carried out   by    a second-order averaging,
 which requires rather tedious mathematical manipulations
  \cite{philkir2,philkir3}. It is also possible to  calculate
 the cross-over  between  the   white noise and  the  Ornstein-Uhlenbeck scaling 
  regimes. The averaging method works at second order  for the 
  Ornstein-Uhlenbeck noise because its  time derivative  is a white noise. However, 
 if the  random excitation is generated from a white noise through a differential
 equation of order $n$, one has  to perform   averaging  at $(n+1)$-th order and 
  in practice, the calculations are intractable. 

 In the present work, we follow an  entirely different  approach to study 
  the  nonlinear oscillator  subject to a parametric Gaussian noise with an
  arbitrary   spectrum 
(with the assumption that the spectrum decays as a power-law at high frequencies).
 We shall use  an effective coarse-grained Markovian description
 of the dynamics, following a technique developed  by   Carmeli and Nitzan \cite{nitzan} 
 (see also \cite{roberts,schimansky} for a similar approach).
  This technique 
  will allow us to calculate analytically  the asymptotic  PDF  which leads to the 
  formulae for the growth of the amplitude, of the momentum
 and the energy transfer. In particular we shall prove that the scaling exponents
 depend  both on  the stiffness of the potential at infinity and on the
 smoothness of the random excitation;  smoother the noise (which corresponds to a
 faster decay of the power spectrum at high frequencies),  less efficient
 is the energy transfer from the bath to the oscillator.  The method used here can be 
 adapted both to additive and to multiplicative noise, and   can also be used
 when  a small friction  is present: the system  reaches at large
 times a non-equilibrium steady state in which physical  observables do not grow
 anymore; the cross-over from power-law growth to this steady state occurs
 when the  rate of energy dissipation by friction matches that of energy absorption
 from the random  environmental loading. 

  The outline of this work is as follows. In Section~\ref{sec:model}, we define
 precisely the model we shall  study. In  Section~\ref{sec:method}, we use
 the underlying integrability of the system to write  exact dynamical equations
 in energy-angle variables and we use  the coarse-grained Markovian
 description  to derive an effective Fokker-Planck equation
 for the energy variable.  In Section~\ref{sec:solutions}, we derive explicit
 formulae for various cases: multiplicative or  additive noise, with or without
 dissipation. This leads to a rather exhaustive description of all the different cases.
 In particular, we verify that this method allows us to recover the analytical
 results obtained previously for white and Ornstein-Uhlenbeck noises.  The last section is devoted
 to concluding remarks. 

\section{The nonlinear oscillator with parametric noise}
\label{sec:model}

  A paradigm for the study of interplay of noise and nonlinearity
 is the   nonlinear oscillator subject to parametric random excitations:
 \begin{equation}
   \frac{\textrm{d}^2 }{\textrm{d} t^2}x(t) + \gamma 
  \frac{\textrm{d} }{\textrm{d} t}x(t)   +(\omega_0^2 + \xi(t))\, x(t) 
   +  \frac { \partial{\mathcal U}(x)}{\partial x}   = 0\,.
 \label{dynamique}
\end{equation}
The variable  $x(t)$ represents  the amplitude of the oscillator
 at time $t$.  The   potential ${\mathcal U}(x)$ that
 confines the  oscillator is assumed to grow 
  faster than quadratically when  $ |x| \rightarrow \infty$ 
  giving rise to nonlinear terms
 in the restoring force.
 We shall make the simplifying assumption that  ${\mathcal U}$
 is an even function of $x$ and behaves 
 as a power-law   of  $x$ when  $ |x| \rightarrow \infty$.  Then,  a suitable rescaling
  of $x$ allows us to write
 \begin{equation}
     {\mathcal U}   \sim \frac{ x^{2\nu}}{2\nu}
  \,\, \hbox{ with } \,\, \nu \ge  2 \,. 
 \label{infU}
\end{equation}
 Typically, $\nu$ is  an integer; the value  $\nu=3$ corresponds to the
 Duffing oscillator.

 The physical interpretation of Eq.~(\ref{dynamique})  is that
  the linear stiffness
 of the  oscillator   fluctuates around its
 mean value $\omega_0^2$ because of randomness in the external
 conditions  and this randomness is represented by 
external noise $\xi(t)$. We also suppose that the oscillator is subject
 to a linear friction with damping coefficient  $\gamma$.

 Equation~(\ref{dynamique})  is thus a nonlinear stochastic differential
 equation.  When the  multiplicative  noise $\xi(t)$
 is a white noise, 
 a coherent  convention to perform stochastic calculus  must be chosen.
   Although Ito calculus is favored by mathematicians, we shall
  use here  the Stratonovich calculus \cite{mortensen,risken}
   which is physically more
  sound because  it   appears  naturally  when one considers the white noise
 as a limit of colored noise with very
  short correlation time  \cite{vankampen,risken}. 
  This equation  seems to be very elementary but it embodies
 many features of random dynamics: inertial effects, nonlinear stiffness
 and parametric noise. In fact, many complex dynamical systems  that
 appear in realistic engineering problems can be reduced after some simplifying
 assumptions to   Equation~(\ref{dynamique}). For example, the torsional
 stability of a suspension bridge under the influence of wind loads can be reduced
 to an equation similar to the one we are studying \cite{ariarat,ibrahim};
 similarly, the dynamics of liquid sloshing,  the  roll motion of a ship  or the stability
 of helicopter  rotor blades in hoover flight  under atmospheric turbulence 
 can be reduced to effective  single-degree-of-freedom systems represented
 by a second order equation with random  parametric vibrations 
 (for explicit derivation of such equations see e.g. \cite{ibrahim}). 
  A similar equation has also been proposed by Fermi as an acceleration
 mechanism for interstellar particles \cite{fermi,bouchet,pollak}.
  Finally, from the mathematical point of view,  Equation~(\ref{dynamique}) 
 is also very appealing: it is  rich  enough to exhibit an interesting
 dynamical  behaviour  but simple enough to allow for explicit solutions 
 \cite{landaMc,hanggirev1,gitterman}.
  This explains why such a simple model  can play  the role of a paradigm.

 The phase space  origin, $x = 0$ and $ \textrm{d}x/\textrm{d}t = 0$
  is a solution
 of  Eq.~(\ref{dynamique}). However, it can be shown that this solution
 is unstable  \cite{bourret,schentzle} when  the power-spectrum 
 of the noise contains all possible frequencies. When  friction is neglected
 (i.e. if the underlying deterministic system is Hamiltonian), then
 because of the  permanent injection of energy into the system by the noise,
 the amplitude, the  velocity and the energy  undergo anomalous diffusion. 
 The associated  anomalous diffusion  exponents and amplitudes have been 
 calculated exactly when the random excitation is a Gaussian white noise 
  \cite{philkir1} or an Ornstein-Uhlenbeck  process \cite{philkir2}.

   In the present work, we   study the effect of
  the  statistical properties of  $\xi(t)$ on the long
 time behaviour of the dynamical variable $x(t)$. 
 We must therefore  specify the  characteristics of the 
 random perturbation   $\xi(t)$.
 We shall  consider  a stationary  Gaussian noise
  of zero mean value. A  Gaussian process is  fully 
 characterized by its auto-correlation function  defined as
 \begin{equation}
  {\mathcal S}(t'-t) = \langle \xi(t')  \xi(t) \rangle  \, .
\label{eq:defS}
 \end{equation}
  In Fourier  space,   the power-spectrum  of the noise is given by 
  \begin{equation}
  \hat{\mathcal S}(\omega) = \int_{-\infty}^{+\infty} {\rm d}t 
        \exp(i\omega t)   {\mathcal S}(t) = 
  \int_{-\infty}^{+\infty} {\rm d}t \exp(i\omega t)
    \langle  \xi(t)  \xi(0)  \rangle  \, .
\label{eq:PSD}
 \end{equation}
  If  $\xi(t)$ is  a  white noise
 of  amplitude ${\mathcal D}$,  we have 
\begin{equation}
  {\mathcal S}(t'-t) =  {\mathcal D} \delta(t'-t)  \quad   \hbox{ and } \quad
   \hat{\mathcal S}(\omega) =   {\mathcal D} \,.
\label{corrwhite}
\end{equation} 
 When $\xi(t)$ is an Ornstein-Uhlenbeck process of amplitude
  ${\mathcal D}$ and  of auto-correlation time $\tau$,   we have 
\begin{equation}
  {\mathcal S}(t'-t) =   
\frac{ {\mathcal D} }{2 \, \tau}   \, {\rm e}^{-|t - t'|/\tau} \quad 
   \hbox{ and } \quad  \hat{\mathcal S}(\omega) =  
   \frac{ {\mathcal D} }{ 1 + \omega^2\tau^2} \, . 
   \label{deftau}
 \end{equation} 
 In this work, we shall consider the case where
the  power-spectrum of $\xi(t)$
   decays  at high frequencies in the following manner
 \begin{equation}
      \hat{\mathcal S}(\omega)  \sim 
    {\mathcal D} (\omega\tau)^{-2\sigma}
   \,\,\,\, \hbox{ when } \,\,\,\,\,  |\omega| \rightarrow \infty \, .
\label{eq:defsigma}
  \end{equation}
  The amplitude  ${\mathcal D}$ of the noise
  and  the correlation-time $\tau$ are defined
   by dimensional analogy with equation~(\ref{deftau}). The 
 exponent $\sigma$ characterizes the high frequency behaviour  of the 
  power-spectrum. When $\sigma$ is an   integer, such a noise 
 can be generated from  the  white noise 
 by solving a linear differential equation of order $\sigma$.

We shall prove that in the long time limit, the statistical properties 
 of the oscillator in the phase space can be   classified
 by the  following  two  parameters 
(i)  the exponent $\nu$, defined in 
  Eq.~(\ref{infU}),  that encodes  the large amplitude behaviour of the
 confining potential ${\mathcal U}$; (ii) the 
  exponent $\sigma$  that   determines  the high frequency behaviour  of 
 the  power-spectrum.  For fixed values of $\nu$ and  $\sigma$, 
 the phase space distribution takes in the long time limit a universal
 form (that also depends on the dimensional  parameters $\gamma$, $\tau$  and 
  ${\mathcal D}$) that we shall calculate.

 \section{Effective dynamics in the asymptotic regime}
 \label{sec:method}

 \subsection{Use of integrability}

 The mechanical energy of the oscillator is defined as
\begin{equation}
  E =  \frac{1}{2}\dot x^2 +  {\mathcal U}(x) \, .
\label{defEnergy}
\end{equation}
 In the absence of noise and dissipation this quantity is conserved.
 This implies that  the Hamiltonian  system  underlying
  Eq.~(\ref{dynamique}) is integrable. It is therefore possible
 to define an action variable $J$ and an  angular  variable $\phi$ 
 so that the transformation $(p=\dot x, x) \to (J,\phi)$ is a canonical
 transformation. 
 For a given value $E$ of the energy, the  angle $\phi$ is given by 
 \cite{landau,lichtenberg}
\begin{eqnarray}
   \phi  = \omega(E) \int_0^ {x}
   \frac{{\textrm d}y}{\sqrt{ 2(E -{\mathcal U}(y))}}  \quad
 \hbox{ with } \quad   \omega(E) = \left( \frac{2}{\pi}  
   \int_0^ {x_{\rm max}}
   \frac{{\textrm d}y}{\sqrt{ 2(E -{\mathcal U}(y))}} \right)^{-1} \, , 
\label{defphi}
\end{eqnarray}
 where $x_{\rm max}$ satisfies $ {\mathcal U}(x_{\rm max}) =E$
  (recall that ${\mathcal U}$
  is an even function).
 With this definition, the range of the  phase $\phi$ is  $2\pi$.
 The action variable $J$ is a function of the energy only
 and is determined by the following equation:
  \begin{equation}
      \frac{\textrm{d}J }{\textrm{d} E} = \frac{1}{\omega(E)} \, .
\label{def:Action}
\end{equation}
  The Hamiltonian equations of motions, in terms of the action-angle variables, 
 read simply:
 \begin{eqnarray}
   \dot J &=& - \frac{\textrm{d}E(J,\phi) }{\textrm{d} \phi} = 0 \\
   \dot \phi  &=& \frac{\textrm{d}E(J,\phi) }{\textrm{d} J} = \omega(E) \, .
\end{eqnarray}
The variables $(E,\phi)$  define a bona fide set of  coordinates on phase space.
 The formulae for  transforming the  variables 
 from position and velocity  to   energy and angle are given by:
 \begin{eqnarray}
 x(E,\phi)  &=&  \sum_{n=-\infty}^{+\infty} x_n(E) {\rm e}^{in\phi} 
\label{Fourierx}  \\
 \dot x (E,\phi)  &=&  \sum_{n=-\infty}^{+\infty} v_n(E) {\rm e}^{in\phi}  =
  i\omega(E) \sum_{n=-\infty}^{+\infty} n x_n(E) {\rm e}^{in\phi} \, . 
\label{Fourierdotx}
\end{eqnarray}
 Here again  to write this  change of variables,  we have used only 
the deterministic and dissipationless parts  of the dynamics. We choose
 the origin of $\phi$ such  that  $x_n(E)$ is a real number
 and that $x_n(E) = x_{-n}(E)$.
 We also write
 \begin{eqnarray}
  \frac {1}{2} x^2(E,\phi) = 
  \sum_{n=-\infty}^{+\infty} y_n(E) {\rm e}^{in\phi}  \, .
 \label{defyn}
\end{eqnarray}
   We then  have
 \begin{eqnarray}
  \frac{\partial }{\partial  E}
 \left(\frac {1}{2} x^2(E,\phi) \right) &=& 
 \sum_{n=-\infty}^{+\infty} \frac{dy_n(E)}{dE}  {\rm e}^{in\phi} \, ,
 \label{eq:dx2dE} \\
  \frac{\textrm{d} }{\textrm{d} t}
 \left(\frac {1}{2} x^2(E,\phi) \right) &=& 
   i\omega(E) \sum_{n=-\infty}^{+\infty} n y_n(E) {\rm e}^{in\phi}  \, .
  \label{eq:dx2dt}
\end{eqnarray}
 In the general case, we find from  Eq.~(\ref{dynamique})
 that the time variation of the energy is given by
\begin{equation}
   \frac{\textrm{d}E }{\textrm{d} t} = -\gamma\dot x^2 + \xi(t)
 \frac{\textrm{d} }{\textrm{d} t}  \left( \frac{x^2}{2} \right) \, .  
\label{eq:Energie}
\end{equation}
   We have used here  the rules of  classical calculus when changing variables 
  \cite{vankampen,risken}: this is allowed because we are working 
  with the Stratonovich interpretation of Eq.~(\ref{dynamique}).
 The energy variation has thus two contributions: a loss term due to
 friction and a stochastic `elastic energy' term due to the work of the
 random  multiplicative force $x\xi(t)$. 

   The time variation of the angle variable is given by
\begin{equation}
  \frac{\textrm{d}\phi }{\textrm{d} t} = \omega(E) +
  \left( \gamma \dot x - x \xi(t) \right)  \omega(E) 
 \frac{\partial x  }{\partial  E} = \omega(E) + 
\gamma \omega(E) \dot x  \frac{\partial x  }{\partial  E}
 -  \xi(t) \omega(E)  \frac{\partial }{\partial  E}
 \left(\frac {1}{2} x^2(E,\phi) \right) \, . 
\label{eq:dynAngle}
\end{equation}

 If we substitute in the two dynamical equations~(\ref{eq:Energie})
  and~(\ref{eq:dynAngle}),
 the expressions given in Eqs.~(\ref{Fourierx}),~(\ref{Fourierdotx}),
 (\ref{eq:dx2dE}) and  (\ref{eq:dx2dt}),   we obtain 
\begin{eqnarray}
    \dot E &=& \gamma \omega^2(E) \sum_{n=-\infty}^{+\infty} 
  \sum_{m=-\infty}^{+\infty} x_n(E)x_m(E) {\rm e}^{i(n+m)\phi}  
   - \xi(t) \omega(E)  \sum_{n=-\infty}^{+\infty}in  y_n(E) {\rm e}^{in\phi}
 \label{eqstoch:E}  \\
    \dot \phi  &=&  \omega(E) +  \gamma  \omega^2(E) \sum_{n=-\infty}^{+\infty} 
    \sum_{m=-\infty}^{+\infty} i n  x_n(E) 
   \frac{dx_m(E)}{dE}  {\rm e}^{i(n+m)\phi} 
 +  \xi(t) \omega(E)  \sum_{n=-\infty}^{+\infty} \frac{dy_n(E)}{dE}
   {\rm e}^{in\phi}    \, . 
\label{eqstoch:Ephi}
\end{eqnarray}
 We emphasize that this  coupled system of stochastic nonlinear equations is rigorously
 equivalent to the initial random dynamical equation~(\ref{dynamique}).

 \subsection{The effective Markovian description}

 Although the problem we study here  is non-Markovian, because the noise has a non-vanishing
 correlation time, it is possible to derive for the associated 
 Probability Distribution Function  $P_t(E,\phi)$ an effective coarse-grained
  Markovian equation using a Kramers-Moyal type expansion \cite{risken,gardiner,vankampen}:
\begin{eqnarray}
 \frac{ \partial P_t(E,\phi)}{\partial t} = \lim_{\delta \to 0^{+}}
\frac{1}{\delta} \sum_{n=1}^\infty  \frac{ (-1)^n}{n!} 
 \sum_{\substack{m+k=n\\m,k \ge 0} } 
 \left(\frac{\partial }{\partial E}\right)^m 
 \left(\frac{\partial }{\partial \phi}\right)^k
 \left[ \,    {\bf M}_{m,k}(E,\phi, t, \delta) 
  P_t(E,\phi)\,  \right]
 \label{eq:KramersMoyal}
\end{eqnarray}
 where we have defined 
\begin{eqnarray}
  {\bf M}_{m,k}(E,\phi, t, \delta)
 &=&  \left\langle (\Delta E_t(\delta))^m (\Delta \phi_t(\delta))^k  \right\rangle \,,
 \\
  \hbox{ with } \quad \Delta E_t(\delta)  &=&   E(t+\delta) - E(t) 
 = \int_0^{\delta}  {\rm d}s  \, \dot E\left( E(t+s),\phi(t+s),t+s\right) \,,
 \label{eq:DelE}  \\
\quad \quad  \hbox{ and } \quad 
 \Delta \phi_t(\delta)   &=&  \phi(t+\delta) -  \phi(t)
  =  \int_0^{\delta} {\rm d}s  \, \dot\phi\left( E(t+s),\phi(t+s),t+s\right) \, . 
  \label{eq:Delphi}  
\end{eqnarray}
 The expressions of $\dot E$ and $\dot \phi$ on the r.h.s. of
 Eqs.~(\ref{eq:DelE}) and~(\ref{eq:Delphi})  are given in 
 Eqs~(\ref{eqstoch:E}) and~(\ref{eqstoch:Ephi}) respectively.

  The time scale $\delta$
 that appears in the Kramers-Moyal expansion
  must be chosen  in a  physically  relevant manner:  $\delta$ has to  be small 
  but must  remain  larger than the intrinsic period of the oscillator (this condition
  is automatically fulfilled  at large amplitudes because the  intrinsic period
  tends to zero).  Besides,  one must also have  $\delta \gg \tau$ 
 (where  $\tau$  characterizes
 the correlation-time of the noise) in order to end up with an effective
 Markovian description of the dynamics. 

  A  systematic procedure for evaluating the coefficients
  ${\bf M}_{m,k}$ that appear in Kramers-Moyal expansion has been developed
by  Carmeli and Nitzan in \cite{nitzan}.
 We first rewrite  Eqs.~(\ref{eq:DelE}) and~(\ref{eq:Delphi})  as 
\begin{eqnarray}
\Delta E_t(\delta)  &=&  \int_0^{\delta}  {\rm d}s  \, 
\dot E\left( E(t)+\Delta E_t(s) ,\phi(t)+\Delta\phi_t(s), t+s \right) \, \\
 \Delta \phi_t(\delta)  &=&  \int_0^{\delta}  {\rm d}s  \, 
\dot \phi\left( E(t)+\Delta E_t(s) ,\phi(t)+\Delta\phi_t(s), t+s \right) \, .
\end{eqnarray}
 The values of   $\Delta E_t(\delta)$ and $\Delta \phi_t(\delta)$ are evaluated
  according to the following iteration scheme, labelled by the integer index $l$
\begin{eqnarray}
\Delta E_t^{(l)}(\delta)  &=&  \int_0^{\delta}  {\rm d}s  \, 
\dot E\left( E(t)+\Delta E_t^{(l-1)}(s) ,\phi(t)+\Delta\phi_t^{(l-1)}(s), t+s \right) \, \\
 \Delta \phi_t^{(l-1)}(\delta)  &=&  \int_0^{\delta}  {\rm d}s  \, 
\dot \phi\left( E(t)+\Delta E_t^{(l-1)}(s) ,\phi(t)+\Delta\phi_t^{(l-1)}(s), t+s \right) \, , 
\end{eqnarray}
 with initial values given by
\begin{eqnarray}
\Delta E_t^{(0)}(s) = 0 \quad \hbox{ and }   \quad 
 \Delta\phi_t^{(0)}(s) = \omega(E) s  \, .
\end{eqnarray}

 Performing this expansion and neglecting
  the terms of order $\delta^n$ with $n>1$, we obtain, in the limit $\delta \to 0$, 
 after some systematic but tedious calculations: 
\begin{eqnarray}
 \frac{1}{\delta} \langle \Delta E(\delta) \rangle &=&  - \gamma \omega^2(E)
     \sum_{n=-\infty}^{+\infty} n^2 |x_n(E)|^2  
  +  \sum_{n=-\infty}^{+\infty} \frac{n^2 \hat{\mathcal S}_n }{4}  
  \left\{  \frac{d \left( \omega(E)y_n(E) \right)^2  }{dE} 
 +  \omega^2(E)
    \frac{d  \left(y_n(E)\right)^2  }{dE} \right\}  \nonumber \\ 
 & &  \quad  \quad {\hskip 4cm} + \frac{\omega^2(E)}{2} 
  \sum_{n=-\infty}^{+\infty} n^2
 |y_n(E)|^2  \frac{d \hat{\mathcal S}_n}{dE}  \, ,  \label{moyE}
 \\
 \frac{1}{2\delta}  \langle (\Delta E)^2(\delta) \rangle &=& \frac{\omega^2(E)}{2}  
   \sum_{n=-\infty}^{+\infty} n^2 |y_n(E)|^2   \hat{\mathcal S}_n 
  \, ,  \label{moyQuadE}   \\ 
\langle \Delta E(\delta) \Delta\phi(\delta)  \rangle &=& 0 \, ,  \label{moyEPhi}
\end{eqnarray}
where we have defined 
 \begin{equation}
 {\mathcal S}_n = {\mathcal S}(n\omega(E)) = 
  \int_{-\infty}^{+\infty} {\rm d}t \exp(i n \omega(E) t)
    \langle  \xi(t)  \xi(0)  \rangle  \, .
\label{def:Sn}
\end{equation}
    We do not need to give  the exact values of  $\langle \Delta\phi(\delta) \rangle$ and 
  $\langle (\Delta\phi)^2(\delta) \rangle$ because they 
 will have no incidence  in the following calculations. Higher moments are negligible
 at the considered order of the calculations.
 
\subsection{Effective Fokker-Planck equation for the energy}

 We now substitute the average values calculated in
  Eqs.~(\ref{moyE},\ref{moyQuadE} and \ref{moyEPhi})
  into the  Kramers-Moyal expansion~(\ref{eq:KramersMoyal}). 
 Because the  cross-correlation term 
 $ \langle \Delta E(\delta) \Delta\phi(\delta)  \rangle$ vanishes,
   we can integrate out the
 angular variable from  Eq.~(\ref{eq:KramersMoyal})  and obtain
 an effective Fokker-Planck equation for the energy:
\begin{equation}
 \frac{ \partial P_t(E)}{\partial t} =  -\frac{\partial }{\partial E}
 \left\{ \,  \frac{\langle \Delta E(\delta) \rangle}{\delta} P_t(E)  \, \right\}
 + \frac{\partial^2}{\partial E^2}
\left\{ \,  \frac{\langle (\Delta E)^2(\delta) \rangle}  {2\delta} P_t(E)   \,\right\} \, .
\label{FPEff1}
\end{equation}
 Defining the following two auxiliary functions:
 \begin{eqnarray}
  \epsilon_1(E) &=&   \sum_{n=-\infty}^{+\infty} n^2 |x_n(E)|^2    \, ,      
   \label{def:eps1} \\
  \epsilon_2(E) &=& \frac{1}{2} 
 \sum_{n=-\infty}^{+\infty} n^2 |y_n(E)|^2   \hat{\mathcal S}_n \, , 
  \label{def:eps2} 
\end{eqnarray}
we rewrite  Eqs.~(\ref{moyE})  and~(\ref{moyQuadE}) as follows:
\begin{eqnarray}
 \frac{1}{\delta} \langle \Delta E(\delta) \rangle &=&   - \gamma \omega^2(E)  \epsilon_1(E)
 +  \omega^2(E)  \frac{{\rm d} \epsilon_2(E) }{ {\rm d} E}  
 +  \epsilon_2(E)  \omega(E)  \frac{{\rm d} \omega(E) }{ {\rm d} E}  \, , \label{moyE2}        \\
 \frac{1}{2\delta}  \langle (\Delta E)^2(\delta) \rangle &=& \omega^2(E) \epsilon_2(E) 
   \, .  \label{moyQuadE2}  
\end{eqnarray}
 Substituting these expressions in Eq.~(\ref{FPEff1}) leads us to  the  effective 
  Fokker-Planck equation for the energy: 
\begin{equation}
 \frac{ \partial P_t(E)}{\partial t} =  \frac{\partial }{\partial E}
 \left\{ \,  \omega(E)
 \left( \gamma\epsilon_1(E) + \epsilon_2(E) \frac{\partial }{\partial E} \right)
\omega(E)P_t(E)  \right\}   \, .
\label{FPEffective}
\end{equation}

  For dissipationless motion, $\gamma=0$, this equation does not have
  a stationary solution: the particle diffuses   in phase space 
  by absorbing energy from the noise 
  and there is no mechanism to limit  the  growth of the amplitude.  The observables 
  grow as power-laws with time as the explicit solutions of the next section
  will show. When  $\gamma \neq 0$,  the system  reaches  a stationary
  measure characterized by a non-equilibrium steady state with an asymptotic
  probability distribution $P_{\hbox{stat}}(E)$ that  differs from the canonical
  Boltzmann-Gibbs law.  For $\gamma\tau \ll 1$,
 the effective Markovian description remains valid and
 the  stationary solution of the effective Fokker-Planck
  equation is  given by
\begin{equation}
 P_{\hbox{stat}}(E) = \frac{{\mathcal N}}{\omega(E)}
 {\hbox{e}}^{-\gamma \int_0^E  {\hbox{d}}u \frac{\epsilon_1(u)}{\epsilon_2(u)} } 
\label{eq:Pstat}
\end{equation}
 where the prefactor ${\mathcal N}$ ensures the normalization of  
 $P_{\hbox{stat}}(E)$.

\section{Explicit solutions} 
\label{sec:solutions}
\subsection{The Hamiltonian  case}

  We shall first consider the case where the  dissipation effects 
  are not taken into account. 
  In the absence of  dissipation,  the  physical observables
  such as the amplitude, the velocity and the energy of the oscillator
  grow  as power-laws with time.  We shall  calculate the associated
  scaling exponents and  prove  that their values    depend  only  on  $\nu$ which 
  determines  the  behaviour of  the   external potential  at large
  amplitudes and on $\sigma$ that  measures the relative weight of 
  high frequencies in  the noise spectrum (and which also 
  characterizes the smoothness of the noise). 

  In the long time limit, the particle  diffuses to large amplitudes in phase
 space. Therefore in Eq.~(\ref{dynamique}) we can neglect the linear restoring
 force (proportional to $\omega_0$) and  replace the potential by
  its asymptotic behaviour given in  Eq.~(\ref{infU}):
 ${\mathcal U}(x)   \sim \frac{ x^{2\nu}}{2\nu} \, .$ Then, the change of variables
 to energy and angle coordinates, given in 
Eqs.~(\ref{defEnergy}) and ~(\ref{defphi}), takes
 the simpler form:
\begin{eqnarray}
   E = \frac{1}{2}\dot x^2 +  \frac{ x^{2\nu}}{2\nu} \,, 
\quad \quad \hbox{ and } \quad 
   \phi  =  \frac{\pi} { 2}
  \frac {  \int_0^ { {x}/{(2\nu E)^{1/{2\nu}}} }
   \frac{{\textrm d}u}{\sqrt{1 -   u^{2\nu}}}}
  {  \int_0^1    \frac{{\textrm d}u}{\sqrt{1 -   u^{2\nu}}}} \,.
 \label{eq:formuleEphi}
\end{eqnarray}
 The equation of motion for the underlying deterministic system
 are given by
 \begin{eqnarray}
   \dot E &=&  0 \\
   \dot \phi  &=&  \omega(E) \quad \hbox{ where } \quad
\omega(E)  =  \frac{\pi} { 2 \sqrt{\nu}}
 \frac { (2\nu E)^{\frac{\nu-1}{2\nu}}}
{  \int_0^1  \frac{{\textrm d}u}{\sqrt{1 -   u^{2\nu}}}}
 =  {\mathcal C}_\nu  E^{\frac{\nu-1}{2\nu}} 
\quad \hbox{ with } \quad
 {\mathcal C}_\nu = \frac{ (2\nu)^{\frac{\nu-1}{2\nu}}\Gamma(\frac{\nu+1}{2\nu}) } 
  {\Gamma(\frac{1}{2\nu})} \sqrt{\pi\nu}\, ,
 \label{eq:formuleomegaE} 
\end{eqnarray}
where the last formula, in terms of the Euler Gamma function, is obtained
from  \cite{abram}.  We  now define \cite{byrd}
 the  hyperelliptic  function ${\mathcal T}_\nu$:
   \begin{equation}
{\mathcal T}_\nu(Y) = X \,  \leftrightarrow 
  Y =  \int_0^X  \frac{{\textrm d}u}{\sqrt{1 -   u^{2\nu}}}  \, .
  \label{hyperelli}
\end{equation}
The  function ${\mathcal T}_\nu$ is periodic with period
\begin{equation}
 t_\nu = 4  \int_0^1  \frac{{\textrm d}u}{\sqrt{1 -   u^{2\nu}}} 
 = \frac{ 2 \sqrt{\pi}} {\nu}  \frac{\Gamma(\frac{1}{2\nu})}
 {\Gamma(\frac{\nu+1}{2\nu})} \, .
\label{def:tnu}
\end{equation} 
 Inverting Eq.~(\ref{eq:formuleEphi}), we express
  the position and the  velocity   in terms
 of   energy and angle using  the  function   ${\mathcal T}_\nu$ \cite{abram,byrd}:
\begin{eqnarray}
          x(E,\phi) &=&  ( 2\nu E)^{\frac{1}{2\nu}} \, 
{\mathcal T}_\nu \left( \frac{ t_\nu \phi}{2\pi}  \right)  \, ,\label{solnx}  \\              
     \dot x(E,\phi) &=& \sqrt{2 E}  \,
  {\mathcal T}_\nu' \left( \frac{ t_\nu \phi}{2\pi}  \right) \, ,
\label{solnv}
 \end{eqnarray}
where ${\mathcal T}_\nu'$ is the derivative of the function 
 ${\mathcal T}_\nu$ which, using Eq.~(\ref{hyperelli}),   satisfies the relation
\begin{equation} 
 \left({\mathcal T}_\nu(Y)\right)^{2\nu}
 +  \left({\mathcal T}_\nu'(Y)\right)^2 = 1  \, .
\label{eq:Idhyp}
\end{equation} 
 The coordinates $x$ and $\dot x$ are $2\pi$ periodic functions of the angle
 variable $\phi$; they can thus be developed into Fourier Series as in 
Eqs.~(\ref{Fourierx}) and~(\ref{Fourierdotx}). More precisely, if we write 
\begin{equation}
 {\mathcal T}_\nu \left( \frac{ t_\nu \phi}{2\pi}  \right) = 
\sum_{n=-\infty}^{+\infty} f_n {\rm e}^{in\phi} \, ,
\end{equation} 
 we obtain  the Fourier coefficients of  $x$ and $\dot x$ 
\begin{equation}
 x_n(E) =  ( 2\nu E)^{\frac{1}{2\nu}} f_n  \quad \quad \hbox{ and }
  \quad   v_n(E) = \sqrt{2 E}\, ( i  n  f_n )  \, .
\end{equation} 
 We note that, in the present case,  the Fourier coefficients depend 
 on the energy $E$  only through a global prefactor that does not depend
 on the harmonic $n$.

  This identification allows to calculate exactly the function
 $\epsilon_1(E)$ defined in Eq.~(\ref{def:eps1}):
\begin{eqnarray} 
  \epsilon_1(E) &=& ( 2\nu E)^{\frac{1}{\nu}}
 \sum_{n=-\infty}^{+\infty} n^2 f_n^2 = \frac{( 2\nu E)^{\frac{1}{\nu}}}{2\pi}
  \left( \frac{ t_\nu }{2\pi}\right)^2      
  \int_0^{2\pi} 
\left|{\mathcal T}_\nu' \left( \frac{ t_\nu \phi}{2\pi}  \right)\right|^2 {\rm d}\phi
 \nonumber \\
  &=&  \frac{( 2\nu E)^{\frac{1}{\nu}}  t_\nu}{(2\pi)^2} \,  4
  \int_0^{ t_{\nu}/4}{\mathcal T}_\nu'(Y)
 \sqrt{ 1 - \left({\mathcal T}_\nu(Y)\right)^{2\nu}}  {\rm d}Y
 =  \frac{( 2\nu E)^{\frac{1}{\nu}}  t_\nu}{\pi^2}  \int_0^1 {\textrm d}u}{\sqrt{1 -   u^{2\nu}}
 \, , \label{calculeps1}
 \end{eqnarray} 
where the second equality is obtained using 
Parseval's identity,  the third equality using Eq.~(\ref{eq:Idhyp}) over a quarter
 of a period of the  hyperellectic function  ${\mathcal T}_\nu$ and the fourth
 equality  by the  change of variable ${\mathcal T}_\nu(Y) = u$.
 Using the expression~(\ref{def:tnu}) and evaluating the last integral in terms of
 Gamma functions \cite{abram}, we obtain:
\begin{equation}
 \epsilon_1(E) =  \frac{( 2\nu E)^{\frac{1}{\nu}} \nu}{\pi \nu( \nu + 1 )}
 \frac{\Gamma^2(\frac{1}{2\nu})}
 {\Gamma^2(\frac{\nu+1}{2\nu})} \, .
\label{eq:formeps1}
 \end{equation}

   We now calculate the function  $\epsilon_2(E)$, defined in Eq.~(\ref{def:eps2}).
 Using Eqs.~(\ref{defyn}) and~(\ref{solnx}), we find that 
 \begin{equation}
 y_n(E) = \frac{ ( 2\nu E)^{\frac{1}{\nu}}}{2} g_n  \quad \quad \hbox{ where } \quad 
   {\mathcal T}_\nu^2 \left( \frac{ t_\nu \phi}{2\pi}  \right) = 
\sum_{n=-\infty}^{+\infty} g_n {\rm e}^{in\phi} \,   \, .
\end{equation} 
 Because   ${\mathcal T}_\nu$ is a real and
 even function of $\phi$, we have $g_n = g_{-n}$ and we can rewrite 
 $\epsilon_2(E)$  as follows:
 \begin{eqnarray}
 \epsilon_2(E) = \frac{ ( 2\nu E)^{\frac{2}{\nu}}}{4}
 \sum_{n=1}^{+\infty} n^2 g_n^2   \hat{\mathcal S}_n \, .
  \label{def2:eps2} 
\end{eqnarray}
 This sum depends in a non-trivial manner on the  noise spectrum. We are, however, interested
 in the long time behaviour of the Probability Distribution Function $\partial P_t(E)$.
 When $t \to \infty$ the typical value of the energy $E$ of the system also
 increases without bounds and, therefore, the intrinsic
 frequency $\omega(E)$ of the system, which according to Eq.~(\ref{eq:formuleomegaE})
  is proportional to $ E^{\frac{\nu-1}{2\nu}}$,  also  increases without bounds.
  Thus, when  $t \to \infty$,  we can replace   $ \hat{\mathcal S}_n 
 ( \, =\hat{\mathcal S}( n\omega(E)) \, )$ by its  asymptotic behaviour given
  in Eq.~(\ref{eq:defsigma}) and   obtain 
 \begin{eqnarray}
 \epsilon_2(E) =  {\mathcal D}  \frac{ ( 2\nu E)^{\frac{2}{\nu}}}
 {4 (\omega(E)\tau)^{2\sigma}}
 \sum_{n=1}^{+\infty} n^{2-2\sigma} g_n^2 \, .
  \label{def3:eps2} 
\end{eqnarray}
 Denoting by  ${\mathcal A}_{\sigma}$ the value of the
 convergent series  $\sum_{n=1}^{+\infty} n^{2-2\sigma} g_n^2$, we can  write
  \begin{equation}
 \epsilon_2(E) = \frac{ {\mathcal D}_{\nu, \sigma}} {\tau^{2\sigma}}
 E^{\frac{2 -\sigma(\nu -1)}{\nu} } \quad \hbox{ with }
  \quad  {\mathcal D}_{\nu, \sigma} = {\mathcal D}  
 \frac{ ( 2\nu)^{\frac{2}{\nu}}}{4 \, ({\mathcal C}_\nu)^{2\sigma}}
{\mathcal A}_{\sigma}  
  \, , 
\label{eq:formegenps2}
 \end{equation}
where ${\mathcal C}_\nu$ was defined in Eq.~(\ref{eq:formuleomegaE}).

  When the noise is white $\sigma =0$,
  its power spectrum is constant and the sum in Eq.~(\ref{def2:eps2})
 can be evaluated exactly. Following steps similar to  those which led to 
  Eq.~(\ref{calculeps1}), we obtain
 \begin{equation}
 \epsilon_2(E) =  {\mathcal D} \frac{( 2\nu E)^{\frac{2}{\nu}}}{4 \pi \nu^2}
 \frac{\Gamma(\frac{1}{2\nu})\Gamma(\frac{3}{2\nu})}
 {\Gamma(\frac{\nu+1}{2\nu})\Gamma(\frac{3\nu+3}{2\nu})} \, .
\label{eq:formeps2}
 \end{equation}

 For an Ornstein-Uhlenbeck noise  $\sigma =1$ and we must evaluate 
 the expression   $\sum_{n=1}^{+\infty}  g_n^2$. This, again can be done
 explicitely, thanks to the Parseval identity:
 \begin{equation}
 \epsilon_2(E) =   {\mathcal D}  \frac{( 2\nu E)^{\frac{3-\nu}{\nu}}}{8\pi \nu \tau^2}
 \left(  \frac{\Gamma(\frac{1}{2\nu})\Gamma(\frac{5}{2\nu})}
 {\Gamma(\frac{\nu+1}{2\nu})\Gamma(\frac{5+\nu}{2\nu})} - 
 \frac{\Gamma^2(\frac{3}{2\nu})}
 {\Gamma^2(\frac{\nu+3}{2\nu})}
 \right)
    \, .
\label{eq:forme2ps2}
 \end{equation}

   We now deduce the asymptotic expression
  of the Probability Distribution Function $\partial P_t(E)$
 in the limit $t \to \infty$ and when there is no dissipation. The effective Fokker-Planck
 equation then reduces to 
\begin{equation}
 \frac{ \partial P_t(E)}{\partial t} =  \frac{\partial }{\partial E}
 \left( \,  \omega(E) \epsilon_2(E) \frac{\partial \, \omega(E)P_t(E) }{\partial E} 
\,  \right) \, .
\label{FPEHamilt}
\end{equation}
 Taking into account the expressions of  $\omega(E)$ and $\epsilon_2(E)$, given in
 Eqs.~(\ref{eq:formuleomegaE}) and~(\ref{eq:formegenps2}) respectively,
 we can rewrite this equation as 
\begin{eqnarray}
 \frac{ \partial P_t(E)}{\partial t} =
  \frac{ {\mathcal D}_{\nu, \sigma} ({\mathcal C}_\nu)^{2}} {\tau^{2\sigma}}
  \frac{\partial }{\partial E}
 \left( \, E^\psi  
  \frac{\partial \, E^{\frac{\nu-1}{2\nu}}  P_t(E) }{\partial E} 
\,  \right) \, \quad \hbox{ with }
   \psi = {\frac{\nu+3 -2\sigma(\nu-1)}{2\nu}}  \, .
\label{FPEHamilt2}
\end{eqnarray}
This equation has a self-similar structure \cite{barenblatt}
 and it is natural to look for solutions  of the form
\begin{equation}
  P_t(E)  = \frac {1}{E}\,\, \phi\left( \frac{ \,\, E^\alpha\,\,}{K t} \right)
 \quad \hbox{ with } \quad  
  K =  \frac{ {\mathcal D}_{\nu, \sigma} ({\mathcal C}_\nu)^2 } {\tau^{2\sigma}} 
  \quad \hbox{ and } \quad 
  \alpha = \frac{(\sigma+1)(\nu-1)}{\nu}   \, , 
\label{eq:defalpha}
\end{equation}
 the prefactor $1/E$  ensures that   $P_t(E)$  is normalized. The function
 $\phi(u)$ of  the scaling
 variable $u = E^\alpha/(Kt)$  satisfies
 an ordinary differential equation.  The solution of this equation  is given by 
\begin{equation}
  \phi(u) \propto u^{\frac{\nu +1}{2\nu\alpha}} {\rm e}^{-\frac{u}{\alpha^2}} \, .
\end{equation}
Inserting this solution into the expression~(\ref{FPEHamilt2}) for  $P_t(E)$
  we find, after normalisation,  the following asymptotic formula for the 
probability distribution function:
\begin{equation}
  P_t(E) = \frac{\alpha}{\Gamma\left(\frac{\nu +1}{2\nu\alpha}\right)}
 \frac{1}{E} 
  \left(\frac{E^\alpha}{\alpha^2 Kt}\right)^{\frac{\nu +1}{2\nu\alpha}}
  \exp\left(-\frac{E^\alpha}{\alpha^2Kt}\right) \, , 
\label{formulePtE}
\end{equation}
 where $K$ and $\alpha$ are defined in  Eq.~(\ref{eq:defalpha}). 

  From  this general result, we can retrieve  the solutions
 for   white noise (W)
 and for  Ornstein-Uhlenbeck  (OU) noise.  For white noise,  using 
 Eq.~(\ref{eq:formeps2})   we have 
\begin{eqnarray}
   { P}_t^{{\rm W}}(E) &=& 
  \frac{1}{  \Gamma \left(\frac{\nu + 1}{2 (\nu-1)}\right)} \,
\frac{\nu-1}{\nu E} \,
\left( \frac{ E^{\frac{\nu-1}{\nu}}}
{2 {\tilde{\mathcal D}}_{{\rm W}}  \, t}\right)^{\frac{\nu+1}{2 (\nu-1)}}
\exp \left\{ - \;\frac{ E^{\frac{\nu-1}{\nu}}}
{2 {\tilde{\mathcal D}}_{{\rm W}} \,  t}
 \right\} \, ,  \nonumber \\
\hbox{ with } \quad   {\tilde{\mathcal D}}_{{\rm W}} &=&  {\mathcal D}
  \frac{( 2\nu )^{\frac{1}{\nu}}(\nu -1)^2 }{2\nu(\nu +1)}
 \frac{\Gamma(\frac{3}{2\nu})\Gamma(\frac{3\nu+1}{2\nu})}
 {\Gamma(\frac{1}{2\nu})\Gamma(\frac{3\nu+3}{2\nu})} \,.
\label{formulePtEwhite}
\end{eqnarray}
 This formula was also obtained  in \cite{philkir1} by  
 stochastic averaging. For  Ornstein-Uhlenbeck   noise, we have,
  using  Eq.~(\ref{eq:forme2ps2})
\begin{eqnarray}
  { P}_t^{{\rm OU}}(E) &=&
  \frac{1}{  \Gamma \left(\frac{\nu + 1}{4 (\nu-1)}\right)} \,
\frac{2(\nu-1)}{\nu E} \,
\left( \frac{ E^{2\frac{\nu-1}{\nu}}}
{2 {\tilde{\mathcal D}}_{{\rm ou}}  \, t}\right)^{\frac{\nu+1}{4 (\nu-1)}}
\exp \left\{ - \;\frac{ E^{2\frac{\nu-1}{\nu}}}
{2 {\tilde{\mathcal D}}_{{\rm ou}} \,  t}
 \right\} \, ,  \nonumber \\
\hbox{ with } \quad   {\tilde{\mathcal D}}_{{\rm ou}} &=& 
 \frac{ {\mathcal D}}{4  \tau^2}  \left( \frac{\nu -1}{\nu}\right)^2 
 ( 2\nu )^{\frac{2}{\nu}}
 \left(  \frac{\Gamma(\frac{\nu+1}{2\nu})\Gamma(\frac{5}{2\nu})}
 {\Gamma(\frac{1}{2\nu})\Gamma(\frac{5+\nu}{2\nu})} - 
 \frac{\Gamma^2(\frac{\nu+1}{2\nu})\Gamma^2(\frac{3}{2\nu})}
 {\Gamma^2(\frac{1}{2\nu})\Gamma^2(\frac{\nu+3}{2\nu})}
 \right) \, .
\label{formulePtEOU}
\end{eqnarray}
 This formula is identical  to the one  derived in \cite{philkir2} by using  
 the  averaging method at the second order.

  To summarize, we have shown that
  in the long time limit  the scaling behaviour of the dissipationless  nonlinear oscillator
  in presence of a noise with a  power-spectrum
  that satisfies equation~(\ref{eq:defsigma}), is given by
  \begin{eqnarray}
     E  &\sim&
    \left(\frac{ {\mathcal D}t} {\tau^{2\sigma}}
   \right)^{\frac{\nu}{(\sigma+1)(\nu-1)}}  \,  , 
   \nonumber   \\
 x  &\sim&  
  \left(\frac{ {\mathcal D}t} { \tau^{2\sigma}}
    \right)^{\frac{1}{2(\sigma+1) (\nu-1)}} \, ,
   \nonumber    \\ 
 \dot x &\sim& 
  \left(\frac{{\mathcal D}t} { \tau^{2\sigma}} 
  \right)^{\frac{\nu}{2(\sigma+1) (\nu-1)}} \,  . 
\label{scalinggeneral}
\end{eqnarray}
These scalings, derived here by a systematic calculation, agree with the
 results given  in \cite{kolkata} that were conjectured  by performing a  partial resummation
 of the small correlation-time expansion of the stochastic Liouville equation
 associated with  the random dynamical system under study. The method used in 
 \cite{philkir2,kolkata} was an approximation  that could not be applied to  the system
 studied here  but only to  a simplified model. In fact  the  resummation
 technique  yielded  the correct exponents but 
  the prefactors were out of reach.  Here, we have obtained 
 the  closed  expression~(\ref{formulePtE})
  for  the  Probability Distribution Function,
 which contains the full information on the statistics
 of the system  when  $t \to \infty$, {\i.e.}, it provides us 
  the  scaling exponents and the corresponding  prefactors.

\subsection{The dissipative case}

  In the presence of dissipation, the system reaches in the long time limit
 a steady  state with the stationary probability given by  Eq.(\ref{eq:Pstat}).
 Using  Eqs.~(\ref{eq:formeps1})  and (\ref{def3:eps2}), we find 
 the  explicit formula  for this stationary distribution:
\begin{equation}
  P_{\rm{stat}}(E) = \frac{\alpha}{\Gamma\left(\frac{\nu +1}{2\nu\alpha}\right)}
 \frac{1}{E} 
  \left(\frac{ 2\nu \gamma \, E^\alpha}{\alpha K (\nu +1)}\right)^{\frac{\nu +1}{2\nu\alpha}}
  \exp\left(-\frac{  2\nu \gamma \, E^\alpha}{\alpha \, K (\nu +1) }\right) \, , 
\label{formulePstationnaire}
\end{equation}
 where $K$ and $\alpha$ are defined in  Eq.~(\ref{eq:defalpha}). 
 From this stationary PDF we find that the typical value of the energy is given by
\begin{equation} 
  E^\alpha \sim \frac{ {\mathcal D}} {\gamma\tau^{2\sigma}} \, .
 \label{eq:Etyp}
 \end{equation}
We observe that the expression of   $P_{\rm{stat}}(E)$
   becomes identical to  that  given  in 
Eq.~(\ref{formulePtE}) if  the time $t$
 in  Eq.~(\ref{formulePtE}) is replaced by $t_\gamma$ with 
\begin{equation}
  t_\gamma = \frac{1}{\gamma} \frac{\nu +1}{2(\sigma +1)(\nu -1)} \, .
\label{def:dissiptime}
\end{equation}
The   value of  $t_\gamma$   determines the time scale at 
  which the system becomes sensitive to the  dissipative effects.
  For $t \ll t_\gamma$, the system evolves as if it were  Hamiltonian
 and the physical observables grow algebraically with time. For 
  $t \gg t_\gamma$, the system sets in its steady state and the statistical averages
 of the physical observables become stationary. Finally, we justify the validity range
 of the stationary probability distribution~(\ref{formulePstationnaire}).
  We recall \cite{nitzan}
  that the Markovian approximation is valid only if $\gamma^{-1}$ is much smaller
 than the  correlation time $\tau$ of the noise, {\it i.e.},
 \begin{equation}
 \gamma \tau \ll 1 \, .
\end{equation}
 Besides, the presence of dissipation should not alter significatively the
 dynamics of the fast variable $\phi$: thus  on the r.h.s. of Eq.~(\ref{eqstoch:Ephi})
  the second term  must remain  much smaller than the first one.
  Using Eqs.~(\ref{eq:formuleomegaE}, \ref{solnx} and \ref{solnv}) this requires that
 $  \gamma  \ll E^{(\nu -1)/2\nu}\,.$
  From  the typical value~(\ref{eq:Etyp}) of the energy
 and the expression of $\alpha$  given in~(\ref{eq:defalpha}), this condition becomes  
 \begin{equation}
 ( \gamma \tau )^{2\sigma}  \gamma^3 \ll  {\mathcal D} \, . 
\end{equation}
  Thus, a sufficient condition is $\gamma^3 <  {\mathcal D}$. For higher values
 of the dissipation rate, the expression~(\ref{formulePstationnaire})
  will no more  be valid. 
  The behaviour of the system can change drastically and a phase transition to a  state
  localized at the origin $x = \dot x = 0$ can occur  \cite{lefever,bourret,kmpmLyap}.

\subsection{The additive noise case}

 In this last subsection, we study 
 the  nonlinear oscillator driven  by an additive noise: 
 \begin{equation}
   \frac{\textrm{d}^2 }{\textrm{d} t^2}x(t) + \gamma 
  \frac{\textrm{d} }{\textrm{d} t}x(t)   
   +  \frac { \partial{\mathcal U}(x)}{\partial x}   =  \xi(t)\,. 
 \label{dynamiqueadditive}
\end{equation}
  The change of coordinates
 to energy and angle variables is  the same as in Eqs.~(\ref{Fourierx})
 and~(\ref{Fourierdotx}). The dynamical equations in  these coordinates
 will  read
\begin{eqnarray}
    \dot E &=& \gamma \omega^2(E) \sum_{n=-\infty}^{+\infty} 
  \sum_{m=-\infty}^{+\infty} x_n(E)x_m(E) {\rm e}^{i(n+m)\phi}  
   + \xi(t) \omega(E)  \sum_{n=-\infty}^{+\infty}in  x_n(E) {\rm e}^{in\phi}
 \label{eqstochadd:E}  \\
    \dot \phi  &=&  \omega(E) +  \gamma  \omega^2(E) \sum_{n=-\infty}^{+\infty} 
    \sum_{m=-\infty}^{+\infty} i n  x_n(E) 
   \frac{dx_m(E)}{dE}  {\rm e}^{i(n+m)\phi} 
 -  \xi(t) \omega(E)  \sum_{n=-\infty}^{+\infty} \frac{dx_n(E)}{dE}
   {\rm e}^{in\phi}   \, .
\label{eqstochadd:Ephi}
\end{eqnarray}
   Using the  Carmeli-Nitzan technique, we derive the following  effective Fokker-Planck
 equation for the energy:
\begin{equation}
 \frac{ \partial P_t(E)}{\partial t} =  \frac{\partial }{\partial E}
 \left\{ \,  \omega(E)
 \left( \gamma\epsilon_1(E) + \tilde\epsilon_2(E) \frac{\partial }{\partial E} \right)
\omega(E)P_t(E)  \right\} \, ,
\label{FPEffectiveadd}
\end{equation}
  the function $\epsilon_1(E)$ was defined in Eq.~(\ref{def:eps1})
 and  $\tilde\epsilon_2(E)$ is given by
 \begin{equation}
  \tilde\epsilon_2(E) = \frac{1}{2} 
 \sum_{n=-\infty}^{+\infty} n^2 |x_n(E)|^2   \hat{\mathcal S}_n \, . 
  \label{def:tildeeps2} 
\end{equation}
 Denoting by  
$\tilde{\mathcal A}_{\sigma}$ the value of the
 convergent series  $\sum_{n=1}^{+\infty} n^{2-2\sigma} f_n^2$, we can  write
  \begin{equation}
  \tilde\epsilon_2(E) = \frac{ \tilde{\mathcal D}_{\nu, \sigma}} {\tau^{2\sigma}}
 E^{\frac{1 -\sigma(\nu -1)}{\nu} } \quad \hbox{ with }
  \quad  \tilde{\mathcal D}_{\nu, \sigma} = {\mathcal D}  
 \frac{ ( 2\nu)^{\frac{2}{\nu}}}{4 \, ({\mathcal C}_\nu)^{2\sigma}}
\tilde{\mathcal A}_{\sigma}  
  \, .
\label{eq:formegenps2tilde}
 \end{equation}
 The constant  ${\mathcal C}_\nu$ was defined in Eq.~(\ref{eq:formuleomegaE}).
It is possible to carry out  explicit calculations  following
 the same lines  as  for the multiplicative noise case.  
 If  dissipation is neglected, 
probability distribution function is found to be 
\begin{equation}
  P_t(E) = \frac{\tilde\alpha}{\Gamma\left(\frac{\nu +1}{2\nu\tilde\alpha}\right)}
 \frac{1}{E} 
  \left(\frac{E^{\tilde\alpha}}
{ {\tilde\alpha}^2 {\tilde K} t}\right)^{\frac{\nu +1}{2\nu\tilde\alpha}}
  \exp\left(-\frac{E^{\tilde\alpha}}{{\tilde\alpha^2}\,{\tilde K}t}\right) \, , 
\label{formulePtEadd}
\end{equation}
 where $\tilde\alpha$ and  $\tilde K$   are given by
\begin{equation}
     \tilde\alpha = \frac{(\sigma+1)(\nu-1) + 1}{\nu}
   \quad \hbox{ and } \quad 
  \tilde K =  \frac{ \tilde{\mathcal D}_{\nu, \sigma} ({\mathcal C}_\nu)^2 }
 {\tau^{2\sigma}}        \,. 
\end{equation}
 For the special cases of white noise or Ornstein-Uhlenbeck noise, explicit
 expressions for $\tilde K$ can be derived and the  formulae for
 the probability distribution function derived in \cite{philkir3} using
 the averaging method are recovered. 

 Thus, in the absence of  dissipation,  the following algebraic scalings
 for the main observables of the system are satisfied
 \begin{eqnarray}
     E  &\sim&
    \left(\frac{ {\mathcal D}t} {\tau^{2\sigma}}
   \right)^{\frac{\nu}{(\sigma+1)(\nu-1) + 1}}  \,  , 
   \nonumber   \\
 x  &\sim&  
  \left(\frac{ {\mathcal D}t} { \tau^{2\sigma}}
    \right)^{\frac{1}{2(\sigma+1)(\nu-1) + 2}  } \, ,
   \nonumber    \\ 
 \dot x &\sim& 
  \left(\frac{{\mathcal D}t} { \tau^{2\sigma}} 
  \right)^{\frac{\nu}{2(\sigma+1)(\nu-1) + 2} } \,  . 
\label{scalinggeneraladd}
\end{eqnarray}
In particular, when the noise is white (or if the potential is quadratic), we
 recover the result  that the energy grows linearly with time.
The amplitude of the noise being constant, the exponents in the additive
 case are smaller than those in the multiplicative case, as expected.

 If we  take dissipation into account, the system reaches a steady
state in the long time limit. The expression of the stationary probability
  matches that  of  Eq.~(\ref{formulePtEadd}) if $t$ is taken to be 
\begin{equation}
  t \rightarrow \frac{1}{\gamma} \frac{\nu +1}
 {2 \left(\nu +\sigma(\nu -1)\right)} \, . 
\label{def:dissiptimeadd}
\end{equation}
 This expression  defines the dissipation time scale  for a nonlinear oscillator subject
 to additive noise.

\section{Conclusion}

  We have used an effective coarse-grained Markovian description to 
  carry out a quantitative analysis 
  of the  nonlinear oscillator confined by a polynomial 
 potential and subject to Gaussian noise of  arbitrary   spectrum 
 that  decays as a power-law at high frequencies. This approach has allowed us to calculate
 the distribution,  in phase space,  of the dynamical system in the long time limit.
 In the absence of dissipation, the particle diffuses without bounds  with an
 anomalous scaling law. The energy transfer from the random perturbation to the particle
 also follows a scaling power-law.  The  diffusion exponents and the corresponding amplitudes
  are  determined exactly. In the presence of dissipation, the system reaches
 a non-equilibrium steady state; the corresponding stationary distribution has also been 
  calculated analytically in the limit of vanishingly small dissipation.
  The advantage of the method used here as compared to stochastic averaging is that
 it can readily be adapted to  any noise spectrum. For white and Ornstein-Uhlenbeck noises,
 the two approaches give identical results.  It would be of great interest to 
 apply this  method to  higher dimensional integrable  systems 
 subject to stochastic  perturbations  such
 as the nonlinear Schr\"odinger equation in a random potential,
  which is often used to study the effect of  nonlinearity on  localization.


\end{document}